\documentclass[12pt]{article}
\usepackage{amssymb}
\usepackage{amsmath}

\usepackage{graphicx}

\ifx\pdfoutput\undefined
    \relax
\else
    \usepackage{epstopdf}
\fi

\makeatletter
\def\numberbysection{\@addtoreset{equation}{section}
 	\def\theequation{\thesection.\arabic{equation}}}
\makeatother

\numberbysection

\newcommand{\be}{\begin{eqnarray}}
\newcommand{\ee}{\end{eqnarray}}
\newcommand{\non}{\nonumber}
\newcommand{\tr}{\mathop{\rm tr}\nolimits}
\newcommand{\csch}{\mathop{\rm csch}\nolimits}
\newcommand{\sech}{\mathop{\rm sech}\nolimits}
\newcommand{\sgn}{\mathop{\rm sgn}\nolimits}

\begin{document}

\begin{titlepage}
\strut\hfill UMTG--243
\vspace{.5in}
\begin{center}

\LARGE Finite size effects in the XXZ and sine-Gordon models\\
\LARGE with two boundaries \\[1.0in]
\large Changrim Ahn \footnote{
       Department of Physics, Ewha Womans University, 
       Seoul 120-750, South Korea}
   and Rafael I. Nepomechie \footnote{
       Physics Department, P.O. Box 248046, University of Miami,
       Coral Gables, FL 33124 USA} \\

\end{center}

\vspace{.5in}

\begin{abstract}
We compute the boundary energy and the Casimir energy for both the
spin-${1\over 2}$ XXZ quantum spin chain and (by means of the
light-cone lattice construction) the massive sine-Gordon model with
both left and right boundaries.  We also derive a nonlinear integral
equation for the ground state of the sine-Gordon model on a finite
interval.  These results, which are based on a recently-proposed Bethe
Ansatz solution, are for {\em general} values of the bulk coupling
constant, and for both diagonal and {\em nondiagonal} boundary
interactions.  However, the boundary parameters are restricted to obey
one complex (two real) constraints.
\end{abstract}
\end{titlepage}

\setcounter{footnote}{0}

\section{Introduction}\label{sec:intro}

The spin-${1\over 2}$ XXZ quantum spin chain and the sine-Gordon
quantum field theory on a finite interval (i.e., with both left and
right boundaries) have applications in statistical mechanics,
condensed matter physics and string theory, and have therefore been
studied intensively, e.g., \cite{Ga} - \cite{CSS2}.  Much of this work
has been restricted to either diagonal boundary interactions \cite{Ga}
- \cite{Sk}, \cite{FS} - \cite{DN} or to special values of the bulk
coupling constant \cite{CSS1, LR, CSS2}, because a solution of the XXZ
chain with general (both diagonal and nondiagonal) boundary terms
\cite{dVGR} has not been available.  A solution of the latter problem
for values of the boundary parameters obeying a linear constraint has
recently been proposed \cite{Ne1, CLSW, Ne2} and confirmed numerically
\cite{NR}.

We exploit here this new solution to compute finite-size corrections
to the ground-state energy of both the XXZ chain and (by means of the
light-cone lattice approach \cite{DDV1, DV, RS}) the massive
sine-Gordon model, in a range of parameter space heretofore not
possible.  In particular, we compute the boundary energy and Casimir
energy, and we derive a Kl\"umper - Pearce - Destri - de Vega
\cite{KP, DDV2} nonlinear integral equation for the ground state of
the sine-Gordon model on a finite interval.

The outline of this article is as follows.  In Section \ref{sec:XXZ},
we consider the open XXZ quantum spin chain with $N$ spins.  After a
brief review of the Bethe Ansatz solution \cite{Ne1, CLSW, Ne2}, we
compute the ground-state energy, in particular the corrections of
order $1$ and $1/N$, and therefore \cite{BCN, Af}, the central charge. 
In Section \ref{sec:SG}, we turn to the sine-Gordon model on an
interval of length $R$.  We observe that this model contains an
additional boundary parameter $|\gamma_{+} - \gamma_{-}|$
which has not previously been noted.  We analyze the light-cone
lattice \cite{LMSS, DDV1, DV, RS} version of this model, which is
formally quite similar to the open XXZ chain.  We determine the
relation between the lattice and continuum boundary parameters by
matching the boundary (order $1$) energies in the corresponding
models.  We then formulate a nonlinear integral equation \cite{KP,
DDV2} for the ground state, and give a corresponding formula for the
Casimir (order $1/R$) energy.  In the ultraviolet $(R \rightarrow 0$)
limit, the central charge of the sine-Gordon model coincides with that
of the XXZ spin chain.  Our result for the Casimir energy at the
free-Fermion point coincides with the result from the TBA approach of
Caux {\it et al.} \cite{CSS2}.  Moreover, we compute the Casimir
energy numerically over a wide range of bulk and boundary parameters,
and track the crossover from the ultraviolet to the infrared regions. 
We conclude in Section \ref{sec:conclude} with a brief discussion of
our results and some interesting open problems.

\section{The open XXZ quantum spin chain}\label{sec:XXZ}

We begin by briefly reviewing the recently-proposed \cite{Ne1, CLSW,
Ne2} Bethe Ansatz solution of the open spin-${1\over 2}$ XXZ quantum
spin chain with both diagonal and nondiagonal boundary terms.  In
terms of the parameters introduced in the latter reference, the 
Hamiltonian is given by
\be
{\cal H }&=& {1\over 2}\Big\{ \sum_{n=1}^{N-1}\left( 
\sigma_{n}^{x}\sigma_{n+1}^{x}+\sigma_{n}^{y}\sigma_{n+1}^{y}
+\cosh \eta\ \sigma_{n}^{z}\sigma_{n+1}^{z}\right)\non \\
&+&\sinh \eta \Big[ 
\coth \alpha_{-} \tanh \beta_{-}\sigma_{1}^{z}
+ \csch \alpha_{-} \sech \beta_{-}\big( 
\cosh \theta_{-}\sigma_{1}^{x} 
+ i\sinh \theta_{-}\sigma_{1}^{y} \big) \non \\
&-& \coth \alpha_{+} \tanh \beta_{+} \sigma_{N}^{z}
+ \csch \alpha_{+} \sech \beta_{+}\big( 
\cosh \theta_{+}\sigma_{N}^{x}
+ i\sinh \theta_{+}\sigma_{N}^{y} \big)
\Big] \Big\} \,,
\label{Hamiltonian}
\ee
where $\sigma^{x} \,, \sigma^{y} \,, \sigma^{z}$ are the usual Pauli
matrices, $\eta$ is the bulk anisotropy parameter, $\alpha_{\pm} \,,
\beta_{\pm} \,, \theta_{\pm}$ are boundary parameters, and $N$
is the number of spins.  
The boundary parameters are assumed to satisfy the linear constraint
\be
\alpha_{-} + \beta_{-} + \alpha_{+} + \beta_{+} = \pm (\theta_{-} - 
\theta_{+}) + \eta k \,,
\label{constraint}
\ee
where $k$ is an even integer if $N$ is odd, and is an odd integer if
$N$ is even. In terms of the ``shifted'' Bethe roots 
$\{ \tilde u_{j}\}$ \cite{NR}, the energy eigenvalues are given by
\be
E &=& \sinh^{2}\eta \sum_{j=1}^{M}{1\over 
\sinh (\tilde u_{j} - {\eta\over 2})\ \sinh(\tilde u_{j}+{\eta\over 2})}
+ {1\over 2} \sinh \eta\left( \coth \alpha_{-} + \tanh \beta_{-} +
\coth \alpha_{+} + \tanh \beta_{+} \right) \non \\
&+& {1\over 2} (N-1) \cosh \eta \,,
\label{energy}
\ee
and the Bethe Ansatz equations are given by
\be
\lefteqn{\left(
{\sinh(\tilde u_{j} + {\eta\over 2}) \over 
\sinh (\tilde u_{j} - {\eta\over 2})}\right)^{2N}
{\sinh(\tilde 2u_{j} + \eta) \over 
\sinh (\tilde 2u_{j} - \eta)}
{\sinh(\tilde u_{j} - {\eta\over 2} + \alpha_{-}) \over 
\sinh (\tilde u_{j} + {\eta\over 2} - \alpha_{-})}
{\cosh(\tilde u_{j} - {\eta\over 2} + \beta_{-}) \over 
\cosh (\tilde u_{j} + {\eta\over 2} - \beta_{-})}} \non \\
& & \times 
{\sinh(\tilde u_{j} - {\eta\over 2} + \alpha_{+}) \over 
\sinh (\tilde u_{j} + {\eta\over 2} - \alpha_{+})}
{\cosh(\tilde u_{j} - {\eta\over 2} + \beta_{+}) \over 
\cosh (\tilde u_{j} + {\eta\over 2} - \beta_{+})}
= - \prod_{k=1}^{M} 
{\sinh(\tilde u_{j} - \tilde u_{k} + \eta) \over 
\sinh (\tilde u_{j} - \tilde u_{k} - \eta)}
{\sinh(\tilde u_{j} + \tilde u_{k} + \eta) \over 
\sinh (\tilde u_{j} + \tilde u_{k} - \eta)} \,, \non \\
& & j = 1 \,, \cdots \,, M \,, 
\label{BAE}
\ee
where the number $M$ of Bethe roots is given by
\be
M={1\over 2}(N-1+k) \,,
\label{Mvalue}
\ee
$k$ being the integer appearing in (\ref{constraint}). The  
case of diagonal boundary terms \cite{ABBBQ, Sk} corresponds
to the limit $\beta_{\pm} \rightarrow \pm \infty$, in which case
the constraint (\ref{constraint}) disappears.

We restrict our attention here to the ``massless'' regime (bulk
anisotropy parameter $\eta$ purely imaginary, with $0 < \Im m\ \eta <
\pi$); and therefore, to ensure Hermiticity of the Hamiltonian
(\ref{Hamiltonian}), we restrict the boundary parameters $\alpha_{\pm}
\,, \theta_{\pm}$ to be purely imaginary, and $\beta_{\pm}$ to be
purely real.  It is convenient to define new bulk and boundary
parameters,
\be
\eta = i \mu \,, \qquad 
\alpha_{\pm} = i \mu a_{\pm} \,, \qquad \beta_{\pm} = \mu b_{\pm} \,,
\qquad \theta_{\pm} = i \mu c_{\pm}
\label{newparams} \,,
\ee 
where $\mu\,, a_{\pm}\,, b_{\pm}\,, c_{\pm}$ are all real, 
with $0 < \mu < \pi$.  We use the periodicity 
$\alpha_{\pm} \mapsto \alpha_{\pm} + 2\pi i$
of the Hamiltonian (\ref{Hamiltonian})  (and in fact, of the full 
transfer matrix) to restrict $\alpha_{\pm}$ to the 
fundamental domain 
$-\pi +{\mu\over 2} < \Im m\ \alpha_{\pm} < \pi +{\mu\over 2}$,
which implies
\be
{1\over 2} -\nu <  a_{\pm}  < {1\over 2} + \nu \,,
\label{ranges}
\ee 
where $\nu \equiv {\pi \over \mu} > 1$. 

Considering separately the imaginary and real parts of
the constraint equation (\ref{constraint}), we see that the boundary
parameters must in fact satisfy a pair of real constraints
\be
a_{-} + a_{+} &=& \pm |c_{-} - c_{+}| + k \,, \non \\
b_{-} + b_{+} &=& 0 \,.
\label{realconstraints}
\ee
The absolute values can be introduced without loss of generality,
since the preceding sign is arbitrary.

The energy eigenvalues $E$ depend on the parameters $c_{\pm}$ only
through the absolute value of their difference, $|c_{-} - c_{+}|$. 
Indeed, by performing on the Hamiltonian (\ref{Hamiltonian}) a global
spin rotation about the $z$ axis, the parameters $c_{\pm}$ are shifted
by a constant, i.e., $c_{\pm} \mapsto c_{\pm} + const$.  In
particular, one can eliminate one of these two parameters (say,
$c_{+}$), which results in a shift of the other ($c_{-} \mapsto c_{-}
- c_{+}$).  Hence, the energy depends on the difference $c_{-} -
c_{+}$.  Furthermore, by performing on the Hamiltonian a time-reversal
(complex-conjugation) transformation, the parameters $c_{\pm}$ are
negated, i.e., $c_{\pm} \mapsto -c_{\pm}$.  Hence, the energy in fact
depends on $|c_{-} - c_{+}|$.

Let us consider the energy of the ground state of this model as a
function of $N$, for large $N$.  The leading (order
$N$) contribution, which does not depend on the boundary interactions,
is well known \cite{YY}.  Our objective here is to compute the boundary
(order $1$) and Casimir (order $1/N$) corrections.

\subsection{Boundary energy}\label{subsec:XXZboundenergy}

Let us streamline the notation by defining the basic quantities
\footnote{We follow the notations used in \cite{DN}.}
\be
e_{n}(\lambda) =
{\sinh \mu \left( \lambda + {i n\over 2} \right) 
\over \sinh \mu \left( \lambda - {i n\over 2} \right) } \,, \qquad
g_{n}(\lambda) = e_{n}(\lambda \pm {i \pi \over 2 \mu})
= {\cosh \mu \left( \lambda + {i n\over 2} \right) 
\over \cosh \mu \left( \lambda - {i n\over 2} \right) } \,.
\ee
The Bethe Ansatz Eqs. (\ref{BAE}) then take the compact form
\be 
e_{1}(\lambda_{j})^{2N+1}\  g_{1}(\lambda_{j})
{e_{2a_{-}-1}(\lambda_{j})\ e_{2a_{+}-1}(\lambda_{j})\over 
g_{1+2ib_{-}}(\lambda_{j})\ g_{1+2ib_{+}}(\lambda_{j})} &=& 
-\prod_{k=1}^M 
e_{2}(\lambda_{j}-\lambda_{k})\ 
e_{2}(\lambda_{j}+\lambda_{k})  \,, \non \\ 
& & j = 1 \,, \cdots \,, M \,,
\label{BAEcompact}
\ee
where we have set $\tilde u_{j} = \mu \lambda_{j}$, and we use the new
parameters introduced in (\ref{newparams}).

\begin{figure}[tb]
	\centering
	\includegraphics[width=0.80\textwidth]{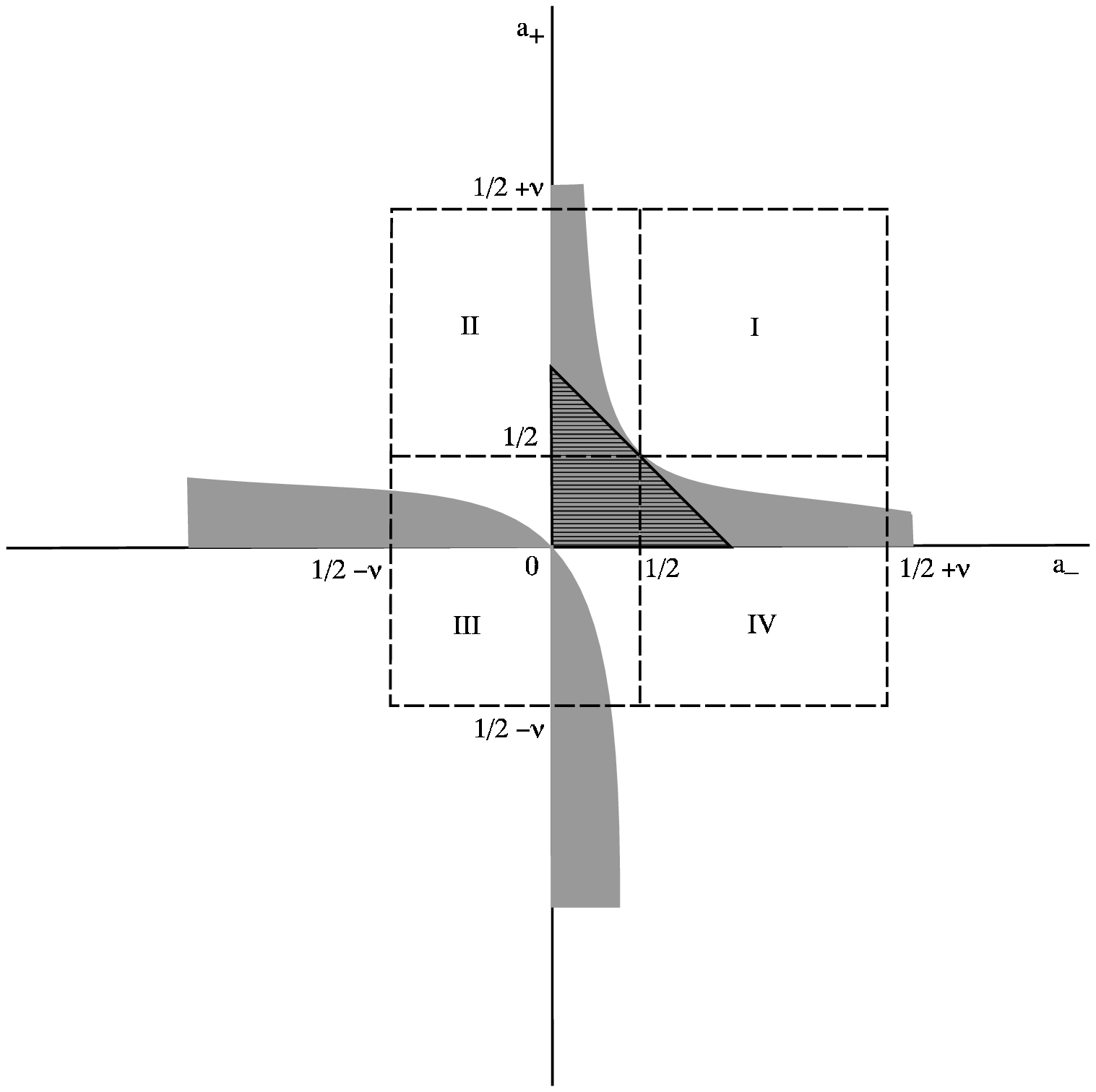}
	\caption[xxx]{\parbox[t]{0.9\textwidth}{
Domain of boundary parameters $a_{\pm}$.  For the ruled triangular region,
the Bethe Ansatz does not give the ground state.  For the shaded regions
(including the special case noted in \cite{NR} corresponding to the
line $a_{+} + a_{-} =1$),
the Bethe Ansatz does give the ground state, but the shifted Bethe roots
are not all real.  For the blank regions (in particular, those labeled I -
IV, as in Eq. (\ref{regions})) the Bethe Ansatz gives the ground state, 
and all roots are real. (Based on numerical results for $N=4$, where 
$b_{\pm}$ and $c_{\pm}$ satisfy Eq. (\ref{realconstraints}) with $k=1$. 
As $N$ increases, the shaded area also increases. We conjecture that for 
$N \rightarrow \infty$, only regions I - IV remain unshaded.)}
	}
	\label{fig:domain}
\end{figure}

We wish to focus here on the {\em ground} state with no holes.  Hence,
we take $N$ even, since states with $N$ odd correspond to excited
states with an odd number of holes.  Moreover, we take (see Eq. 
(\ref{Mvalue}))
\be
k=1 \,, \qquad M ={N\over 2} \,.
\label{kvalue}
\ee
According to \cite{NR}, for this case the Bethe Ansatz solution
correctly yields the energy of the ground state, and the shifted Bethe
roots corresponding to this state are real.
However, we have subsequently found through further numerical studies
of chains with small values of $N$ that this statement must be
qualified: there are regions in the parameter space (\ref{ranges}) for
which some of the shifted Bethe roots are imaginary (presumably
corresponding to boundary bound states), or for which the Bethe Ansatz
does not yield the ground state.  (See Figure  \ref{fig:domain}.)  
For simplicity, we henceforth restrict the boundary parameters
$a_{\pm}$ to the following four regions,
\be
\begin{array}{r@{\ : \quad}l}
    I   & {1\over 2} < a_{\pm} < {1\over 2} + \nu \\
    II  & {1\over 2} < a_{+} < {1\over 2} + \nu \quad \& \quad 
    {1\over 2} - \nu < a_{-} < 0  \\
    III & {1\over 2} - \nu < a_{\pm} < 0 \\
    IV  & {1\over 2} - \nu < a_{+} < 0 \quad \& \quad 
    {1\over 2} < a_{-} < {1\over 2} + \nu
  \end{array}
  \label{regions}
\ee
for which our numerical results indicate that the Bethe Ansatz
solution does yield the energy of the ground state, and the shifted
Bethe roots corresponding to this state are all real.

We remark that (\ref{ranges}) implies that $-2\nu < a_{-} + a_{+} - 1
< 2\nu$; and hence the first constraint in Eq. 
(\ref{realconstraints}) with $k=1$ implies
\be
|c_{-} - c_{+}| < 2\nu \,.
\label{cdiff}
\ee
This condition can always be satisfied, since the Hamiltonian and
transfer matrix also have the periodicity $\theta_{\pm} \mapsto
\theta_{\pm} + 2\pi i$, which corresponds to $c_{\pm} \mapsto c_{\pm}
+ 2\nu$.

In order to compute the energy of the ground state for large $N$, we
first determine the density of (real) Bethe roots for this state.  To
this end, we take the logarithm of the Bethe Ansatz Eqs. 
(\ref{BAEcompact}) and obtain
\be
h(\lambda_{j}) = j \,, \qquad j = 1 \,, \cdots \,, {N\over 2} \,,
\label{BAElog}
\ee
where the counting function $h(\lambda)$ is given by
\be
h(\lambda)&=&{1\over 2\pi}\Big\{ (2N+1) q_{1}(\lambda) + r_{1}(\lambda)
+ q_{2a_{-}-1}(\lambda) -  r_{1+2ib_{-}}(\lambda)
+ q_{2a_{+}-1}(\lambda) -  r_{1+2ib_{+}}(\lambda) \non \\
&-& \sum_{k=1}^{N\over 2} \left[ q_{2}(\lambda - \lambda_{k}) +
q_{2}(\lambda + \lambda_{k})  \right] \Big\} \,,
\label{counting}
\ee
where $q_{n}(\lambda)$ and $r_{n}(\lambda)$ are odd functions defined
by
\be
q_{n}(\lambda) &=& \pi + i \ln e_{n}(\lambda) 
= 2 \tan^{-1}\left( \cot(n \mu/ 2) \tanh( \mu \lambda) \right)
\,, \non \\
r_{n}(\lambda) &=&  i \ln g_{n}(\lambda) \,.
\label{logfuncts}
\ee
We have checked numerically that, for the ground state, the 
right-hand-side of (\ref{BAElog}) is indeed given by successive integers 
from $1$ to $N/2$ \cite{Ga, ABBBQ}.
The Bethe roots $\{ \lambda_{k} \}$ can all be chosen to be strictly
positive. Then, defining $\lambda_{-k} \equiv -\lambda_{k}$, we  
rewrite the last term in (\ref{counting}) more symmetrically as
follows:
\be
-\sum_{k=1}^{N\over 2} \left[ q_{2}(\lambda - \lambda_{k}) +
q_{2}(\lambda + \lambda_{k})  \right] =
-\sum_{k=-{N\over 2}}^{N\over 2} q_{2}(\lambda - \lambda_{k}) 
+ q_{2}(\lambda) \,.
\ee
The root density $\rho(\lambda)$ for the ground state is 
therefore given by
\be
\rho(\lambda) &=& {1\over N} {d h\over d\lambda} \non \\
 &=& 2 a_{1}(\lambda)
 - \int_{-\infty}^{\infty} d\lambda'\ a_{2}(\lambda - \lambda')\
 \rho(\lambda') \label{rhointegraleqn} \\ 
 &+& {1\over N} \left[ a_{1}(\lambda) + b_{1}(\lambda) + a_{2}(\lambda) 
+ a_{2a_{-}-1}(\lambda) -  b_{1+2ib_{-}}(\lambda)
+ a_{2a_{+}-1}(\lambda) -  b_{1+2ib_{+}}(\lambda) \right] \,, \non 
\ee
where we have ignored corrections of higher order in $1/N$ when
passing from a sum to an integral, and we have introduced the
notations
\be
a_n(\lambda) &=& {1\over 2\pi} {d \over d\lambda} q_n (\lambda)
= {\mu \over \pi} 
{\sin (n \mu)\over \cosh(2 \mu \lambda) - \cos (n \mu)} \,, \non \\
b_n(\lambda) &=& {1\over 2\pi} {d \over d\lambda} r_n (\lambda)
= -{\mu \over \pi} 
{\sin (n \mu)\over \cosh(2 \mu \lambda) + \cos (n \mu)} \,. 
\ee 
The linear integral equation (\ref{rhointegraleqn}) for $\rho(\lambda)$
is readily solved by Fourier transforms,
\be
\rho(\lambda) = 2 s(\lambda) + {1\over N} R(\lambda) \,,
\label{rho}
\ee
where
\be
s(\lambda) &=& {1\over 2\pi} \int_{-\infty}^{\infty} d\omega\ 
e^{-i \omega \lambda} {1\over 2 \cosh(\omega/2)} 
= {1\over 2 \cosh (\pi \lambda)} \,, \label{kernels} \\
R(\lambda) &=& {1\over 2\pi} \int_{-\infty}^{\infty} d\omega\ 
e^{-i \omega \lambda} \Big\{ 
{\sinh((\nu-2)\omega/4) \cosh(\nu \omega/4)\over 
\sinh((\nu-1)\omega/2) \cosh(\omega/2)} 
+ {\sinh((\nu-2)\omega/2) \over 
2\sinh((\nu-1)\omega/2) \cosh(\omega/2)} \non \\
&+&  \sgn(2a_{+}-1){\sinh((\nu-|2a_{+}-1|)\omega/2) \over 
2\sinh((\nu-1)\omega/2) \cosh(\omega/2)} + 
{\sinh((1+2ib_{+})\omega/2)\over 
2\sinh((\nu-1)\omega/2) \cosh(\omega/2)} + (+ \leftrightarrow -)
\Big\} \,, \non 
\ee 
which we have obtained using the results \footnote{Our 
conventions are
\be
\hat f(\omega) \equiv \int_{-\infty}^\infty e^{i \omega \lambda}\ 
f(\lambda)\ d\lambda \,, \qquad\qquad
f(\lambda) = {1\over 2\pi} \int_{-\infty}^\infty e^{-i \omega \lambda}\ 
\hat f(\omega)\ d\omega \,. \non 
\ee} 
\be
\hat a_{n}(\omega) &=& \sgn(n) {\sinh \left( (\nu  - |n|) 
\omega / 2 \right) \over
\sinh \left( \nu \omega / 2 \right)} \,,
\qquad 0 \le |n| < 2 \nu  \,, \label{fourier1} \\
\hat b_{n}(\omega) &=&
-{\sinh \left( n \omega / 2 \right) \over
\sinh \left( \nu \omega / 2 \right)} \,,
\qquad \qquad\qquad\quad  0 < \Re e\ n < \nu  \,, 
\label{fourier2}
\ee
where $\nu = {\pi \over \mu} > 1$, and the sign function 
$\sgn(n)$ is defined as 
\be
\sgn(n) = \left\{ \begin{array}{c@{\quad : \quad} l}
{n\over |n|} & n \ne 0 \\
0 & n=0
\end{array} \right. \,.
\ee
We have also made use of the fact $|2 a_{\pm} - 1| < 2\nu$, which
follows from (\ref{ranges}).

Having determined the root density for the ground state up to order
$1/N$, we now proceed to compute the corresponding energy.
Recalling the result (\ref{energy}) for the energy in terms of the 
Bethe roots, we obtain 
\be
E &=& - {2\pi \sin \mu\over \mu} \sum_{j=1}^{N\over 2} 
a_{1}(\lambda_{j}) + \ldots
= - {\pi \sin \mu\over \mu}\left[ 
\sum_{j=-{N\over 2}}^{N\over 2} 
 a_{1}(\lambda_{j}) - a_{1}(0) \right] + \ldots \non \\
&=& - {\pi \sin \mu\over \mu} \left[ 
N\int_{-\infty}^{\infty}d\lambda\ a_{1}(\lambda)\ \rho(\lambda)
- a_{1}(0) \right] + \ldots \,,
\ee
where again we ignore corrections that are higher order in $1/N$,
and the ellipsis $(\ldots)$ denotes the terms in (\ref{energy})
which do not depend on the Bethe roots.
Substituting the result (\ref{rho}) for the root density, we 
arrive at the final result for the ground-state energy
\be
E = E_{bulk} + E_{boundary} \,,
\label{XXZtotalenergy}
\ee
where
\be
E_{bulk} &=& - {2N \pi \sin \mu\over \mu} 
\int_{-\infty}^{\infty}d\lambda\ a_{1}(\lambda)\ s(\lambda) 
+ {1\over 2}N \cos \mu 
\non \\
 &=&  - N \sin^{2} \mu \int_{-\infty}^{\infty}
 d\lambda\ {1\over \left(\cosh(2 \mu \lambda) - \cos \mu \right) 
\cosh (\pi \lambda)} +  {1\over 2}N \cos \mu \,,
\label{XXZbulkenergy}
\ee
which is the well-known \cite{YY} result for the bulk (order $N$)
ground-state energy of the XXZ chain; and the boundary (order $1$)
energy is given by
\be
E_{boundary} &=& - {\pi \sin \mu\over \mu} 
\int_{-\infty}^{\infty}d\lambda\ a_{1}(\lambda) \left[
R(\lambda) - \delta(\lambda) \right] \non \\
&+& {1\over 2} \sin \mu \left( \cot \mu a_{-} + i\tanh \mu b_{-} +
\cot \mu a_{+} + i\tanh \mu b_{+} \right) -{1\over 2}\cos \mu
\,,
\label{XXZboundenergy}
\ee
where $R(\lambda)$ is given by (\ref{kernels}).  It should be
understood that the boundary parameters obey the constraints
(\ref{realconstraints}) with $k=1$.  In the limit of diagonal boundary
terms $b_{\pm} \rightarrow \pm \infty$, this result
for the boundary energy agrees with that of \cite{HQB}. 

The result (\ref{XXZboundenergy}) for the boundary energy, which is
the sum of contributions from both boundaries, implies that the
contribution of each boundary is given by
\be
E_{boundary}^{\pm} &=& - {\sin \mu\over 2\mu} 
\int_{-\infty}^{\infty} d\omega\ 
{1\over 2\cosh (\omega/ 2)}
\Big\{ 
{\sinh((\nu-2)\omega/4) \over 2\sinh(\nu \omega/4)} 
-{1\over 2} \non \\
&+& \sgn(2a_{\pm}-1){\sinh((\nu-|2a_{\pm}-1|)\omega/2) \over 
2\sinh((\nu-1)\omega/2) \cosh(\omega/2)} + 
{\sinh(\omega/2) \cos (b_{\pm}\omega) \over 
\sinh(\nu \omega/2)} \Big\} \non \\
&+& {1\over 2} \sin \mu \cot \mu a_{\pm} 
 -{1\over 4}\cos \mu \,.
\label{XXZboundenergyeach}
\ee
Indeed, as already noted, the total energy depends on $c_{\pm}$ only
through the combination $|c_{-}-c_{+}|$.  Hence, the left and right
boundary energies must be independent of $c_{\pm}$.  \footnote{For
example, consider the right boundary energy $E_{boundary}^{+}$.  If it
does depend on $c_{+}$, then it must also depend on $c_{-}$, since the
dependence must be of the form $|c_{-}-c_{+}|$.  But the right
boundary energy can depend only on the right boundary parameters. 
Hence, it cannot depend on the left boundary parameter $c_{-}$; and
therefore, it cannot depend on $c_{+}$.}
Let us now consider the left-right symmetric boundary case, with
$a_{+}= a_{-}$ and $b_{+}= -b_{-}$ (and $c_{\pm}$ are arbitrary, so
that $a_{+}= a_{-}$ is arbitrary).  For this case, we expect that the
energy contributions of the left and right boundaries are equal,
$E_{boundary}^{-} = E_{boundary}^{+}$.  Dividing
(\ref{XXZboundenergy}) in half, we obtain the result
(\ref{XXZboundenergyeach}).  We now argue that this result holds for
the general (nonsymmetric) case.  First, the form of
the Hamiltonian (\ref{Hamiltonian}) implies that the functional
dependence of the right boundary energy on the right boundary
parameters should be the same as the functional dependence of the left
boundary energy on the left boundary parameters; i.e.,
$E_{boundary}^{+}=f(a_{+}\,, b_{+})$ and $E_{boundary}^{-}=f(a_{-}\,,
b_{-})$ with the same function $f$.  The boundary energy expressions
(\ref{XXZboundenergyeach}) evidently satisfy this property.  Finally,
the left and right boundary energies must be independent of each
other.  Hence, having computed $E_{boundary}^{+}(a_{+}\,, b_{+})$ for
the left-right symmetric boundary case for arbitrary $a_{+}$ and
$b_{+}$, it cannot change if we change $a_{-}$ and/or $b_{-}$. 
(Although we do not know the Bethe Ansatz when $b_{+} \ne - b_{-}$,
one could in principle do the computations numerically.)  Thus, the
expression $E_{boundary}^{+}(a_{+}\,, b_{+})$ must be correct even for
the left-right nonsymmetric case.  Similarly,
$E_{boundary}^{-}(a_{-}\,, b_{-})$ must be correct even for the
left-right nonsymmetric case.

\subsection{Casimir energy}\label{subsec:XXZcasimirenergy}

The computation of the Casimir (order $1/N$) energy requires
considerably more effort.  A systematic approach based on the
Euler-Maclaurin formula \cite{WW} and Wiener-Hopf integral equations
\cite{YY} was developed in \cite{WE} for the periodic XXZ chain, and
was extended to the open XXZ chain with diagonal boundary terms in
\cite{HQB}.  Fortunately, the analysis of our system of Bethe Ansatz
equations (\ref{energy}), (\ref{BAE}) is very similar to the one
presented in \cite{HQB}.  Hence, we shall indicate only the
significant differences which occur with respect to this reference,
which we now denote by I. Using the fact that 
$q_{n}(\lambda) \rightarrow \sgn(n)\pi - \mu n$ 
for $\lambda \rightarrow \infty$, we find that
the ``sum rule'' (I3.30) becomes
\be
\int_{\Lambda}^{\infty} d\lambda\ \rho(\lambda) = {1\over N} 
\left[{1\over 2}\left(1 + s_{-} + s_{+}\right) 
- {1\over \nu} \left( a_{-}  + a_{+}
- ib_{-} -ib_{+}-1 \right) \right] \,,
\label{sumrule}
\ee
where $s_{\pm} \equiv \sgn(a_{\pm} - {1\over 2})$;
and the quantity $\alpha$ (I3.34) becomes
\be
\alpha = {1\over G_{+}(0)}
\left[{1\over 2}\left(s_{-} + s_{+}\right)  
- {1\over \nu}\left( a_{-}  + a_{+}
- ib_{-} -ib_{+}-1 \right) \right] 
\,,
\ee
where, in our conventions, $G_{+}(0)^{2} = 2(\nu -1)/\nu$.
The Casimir energy (i.e., the contribution of order $1/N$ to the 
ground-state energy) is given by 
\be
E_{Casimir} = -{c \pi^{2} \sin \mu\over 24 \mu N} \,,
\label{XXZCasimir}
\ee
where \cite{BCN, Af} the central charge $c$ is given by (I3.38) 
\footnote{In the diagonal limit, the corresponding result is 
\be
c = 1 - {6\over \nu (\nu -1)}\left[
{\nu\over 2}\left(s_{-} + s_{+}\right)  
- \left( a_{-}  + a_{+} \right) \right]^{2} 
\,. \non
\ee
In particular, the central charge equals 1 for the case
$a_{+}+a_{-}=0$ where the boundary fields are real and opposite
\cite{BS}, as well as for the case $a_{\pm}= \nu/2$ of vanishing
boundary fields. In \cite{HQB}, it is implicitly assumed that 
$a_{\pm} > 1/2$, in which case $s_{\pm}=1$.}
\be
c &=& 1 - 12 \alpha^{2} \non \\
&=&  1 - {6\over \nu (\nu -1)}\left[
{\nu\over 2}\left(s_{-} + s_{+}\right)  
- \left( a_{-}  + a_{+}
- ib_{-} -ib_{+}-1 \right) \right]^{2}  \,.
\label{centralcharge1}
\ee
Imposing the constraints (\ref{realconstraints}) with $k=1$
gives the final result 
\be
c = 1 - {6\over \nu (\nu -1)}\left[
{\nu\over 2}\left(s_{-} + s_{+}\right)  
\mp |c_{-} - c_{+}| \right]^{2} \,.
\label{centralcharge2}
\ee

Since the root density should be nonnegative, it follows from the sum
rule (\ref{sumrule}) that the boundary parameters should obey
\be
{1\over 2}\left(1 + s_{-} + s_{+}\right) 
- {1\over \nu} \left( a_{-}  + a_{+}
- ib_{-} -ib_{+}-1 \right) =
{1\over 2}\left(1 + s_{-} + s_{+}\right) 
\mp {1\over \nu}|c_{-} - c_{+}| \ge 0\,,
\ee
i.e., $\pm |c_{-} - c_{+}| \le \nu (1 + s_{-} + s_{+})/2$, which is a
further restriction of the constraint (\ref{cdiff}).

\section{The sine-Gordon model with two boundaries}\label{sec:SG}

We turn now to the sine-Gordon quantum field theory on the finite
``spatial'' interval $x \in \left[ x_{-} \,, x_{+} \right]$, with Euclidean
action
\be
{\cal A} = \int_{-\infty}^{\infty}dy 
\int_{x_{-}}^{x_{+}}dx\  A(\varphi \,, \partial_{\mu} \varphi) 
+ \int_{-\infty}^{\infty}dy \left[ 
B_{-}(\varphi \,, {d\varphi\over dy} )\Big\vert_{x=x_{-}} +
B_{+}(\varphi \,, {d\varphi\over dy} )\Big\vert_{x=x_{+}} \right] \,,
\label{SGaction}
\ee 
where the bulk action is given by
\be 
A(\varphi \,, \partial_{\mu} \varphi) = 
{1\over 2}(\partial_{\mu} \varphi)^{2}
+ \mu_{bulk} \cos (\beta \varphi) \,,
\label{SGbulkaction}
\ee 
and the boundary action is given by 
\be
B_{\pm}(\varphi \,, {d\varphi\over dy} ) = 
\mu_{\pm} \cos( {\beta\over 2} (\varphi - \varphi_{0}^{\pm}))
\pm {\pi\gamma_{\pm}\over \beta} {d\varphi\over dy}  \,.
\label{SGboundaction}
\ee
This action is similar to the one considered by Ghoshal and
Zamolodchikov \cite{GZ}, except that now there are two boundaries
instead of one, and the boundary action (\ref{SGboundaction}) contains
an additional term depending on the ``time'' derivative of the field. 
In the one-boundary case, such a term can be eliminated by adding to
the bulk action (\ref{SGbulkaction}) a term proportional to
$\partial_{x} \partial_{y} \varphi$, which has no effect on the 
classical equations of motion.  However, in the two-boundary
case, one can eliminate in this way only one of the two 
$\gamma_{\pm}$
parameters (say, $\gamma_{+}$), which results in a shift of
the other ($\gamma_{-} \mapsto \gamma_{-} - \gamma_{+}$).
Notice that this discussion is completely parallel to the one 
for the $\theta_{\pm}$ parameters of the open XXZ chain 
(\ref{Hamiltonian}). Indeed, we shall argue below that these two sets 
of parameters are related (\ref{thetareltn}).

Let us consider the energy of the ground state of this model as a
function of the interval length $R \equiv x_{+} - x_{-}$, for large
$R$.  The leading (order $R$) contribution, which does not depend on
the boundary interactions, is well known \cite{AZ2}.  The boundary
(order $1$) correction is also known \cite{AZ1, BPT}.  Our main
objective here is to compute the Casimir (order $1/R$) correction, and
to derive a nonlinear integral equation \cite{KP, DDV2, FMQR} for the
ground state.  We proceed, following the analysis \cite{LMSS} of the
case of Dirichlet boundary conditions, by considering the light-cone
lattice \cite{DDV1, DV, RS} version of this model, defined on a
lattice with spacing $\Delta$.  This lattice model is formally quite
similar to the XXZ chain considered in the previous Section, the main
difference being the introduction of an alternating inhomogeneity
parameter $\pm\Lambda$.  The continuum limit consists of taking $N
\rightarrow \infty$, $\Delta \rightarrow 0$, and $\Lambda \rightarrow
\infty$, such that the length $R$ and the soliton mass $m$ (whose
relation to $\mu_{bulk}$ is known \cite{AZ2}) are given by
\be
R = N \Delta \,, \qquad m={2\over \Delta} e^{-\pi \Lambda} \,,
\label{continuumlimit}
\ee
respectively.

In this approach it is evidently necessary to know the relation
between all the parameters of the lattice model and those of the
continuum quantum field theory (\ref{SGaction}) -
(\ref{SGboundaction}).  The relation between the lattice and
continuum bulk coupling constants is well known 
(see, e.g., \cite{FS, DN}):
$\beta^{2} = 8(\pi - \mu) = 8 \pi (\nu - 1)/\nu$, and therefore 
\footnote{It should be clear from the context whether the symbol
$\lambda$ refers to the value (\ref{bulkparamreltn}) of the bulk
coupling constant or to the rapidity variable, as in
(\ref{SGcounting}).  Also, we note that in \cite{DN}, the bulk
coupling constant $\mu$ is restricted to the range $0 < \mu <
{\pi\over 2}$, and the Hamiltonian has a coefficient $\epsilon$, so
that the ``repulsive'' and ``attractive'' regimes correspond to
$\epsilon=+1$ and $\epsilon=-1$, respectively.  Here we instead let
$\mu$ have an increased range $0 < \mu < \pi$, and consider a single
sign of the Hamiltonian, corresponding to the {\em repulsive} regime
in \cite{DN}.  Thus, here the repulsive and attractive regimes
correspond to the ranges $0 < \mu < {\pi\over 2}$ and ${\pi\over 2} <
\mu < \pi$, respectively. In terms of $\nu = \pi/\mu$, these ranges are
$\nu >2$ and $1 < \nu <2$, respectively.}
\be
\lambda \equiv {8\pi\over \beta^{2}} - 1 = {1\over \nu -1} \,.
\label{bulkparamreltn}
\ee
However, corresponding relations for the {\em boundary} parameters
have been known only for the special case of Dirichlet boundary
conditions \cite{LMSS}.

We determine the general relation between the lattice and continuum
boundary parameters in Section \ref{subsec:SGboundenergy} by matching
the boundary energies in the lattice and continuum models.  We then
formulate a nonlinear integral equation for the ground state, and give
a corresponding formula for the Casimir energy.  We examine the
ultraviolet $(R \rightarrow 0$) limit, and also compare our result at
the free-Fermion point ($\lambda=1$) with that of the TBA approach
\cite{CSS2}.  Moreover, we compute the Casimir energy numerically over
a wide range of bulk and boundary parameters, and track the crossover
from the ultraviolet to the infrared regions.

\subsection{Boundary energy and boundary parameters}
\label{subsec:SGboundenergy}

For the light-cone lattice model, the Bethe Ansatz equations can again
be written in the logarithmic form (\ref{BAElog}), except that the
counting function is now given by
\be
h(\lambda)&=&{1\over 2\pi}\Big\{ N \left[ q_{1}(\lambda + \Lambda)
+q_{1}(\lambda - \Lambda) \right]
+q_{1}(\lambda) + r_{1}(\lambda)
+ q_{2a_{-}-1}(\lambda) -  r_{1+2ib_{-}}(\lambda) \non \\
&+& q_{2a_{+}-1}(\lambda) -  r_{1+2ib_{+}}(\lambda)
- \sum_{k=1}^{N\over 2} \left[ q_{2}(\lambda - \lambda_{k}) +
q_{2}(\lambda + \lambda_{k})  \right] \Big\} \,,
\label{SGcounting}
\ee
which depends on the inhomogeneity parameter $\Lambda$. 

The computation of the ground-state root density to order $1/N$
proceeds as in Section \ref{subsec:XXZboundenergy}, and we obtain
\be
\rho(\lambda) = s(\lambda + \Lambda) +  s(\lambda - \Lambda) 
+ {1\over N} R(\lambda) \,,
\label{SGrho}
\ee
where $s(\lambda)$ and $R(\lambda)$ are given by (\ref{kernels}).
Moreover, following \cite{LMSS, RS}, the energy is given by\footnote{We 
consider explicitly here only contributions to the energy which depend on
the Bethe roots.}
\be
E &=& - {1\over \Delta} \sum_{j=1}^{N\over 2} 
\left[ a_{1}(\lambda_{j}+ \Lambda) + a_{1}(\lambda_{j} - \Lambda) \right] 
\non \\
&=& - {1\over \Delta} \left\{
N\int_{-\infty}^{\infty}d\lambda\ a_{1}(\Lambda - \lambda) 
\rho(\lambda) - a_{1}(\Lambda) \right\}  \,,
\ee
where $\Delta$ is the lattice spacing. Substituting the result
(\ref{SGrho}) for the root density, we obtain
\be
E = E_{bulk} + E_{boundary} \,,
\label{SGtotalenergy}
\ee
where
\be
E_{bulk} &=& - {N\over \Delta} 
\int_{-\infty}^{\infty}d\lambda\ a_{1}(\Lambda - \lambda)
\left[ s(\lambda + \Lambda) +  s(\lambda - \Lambda) \right] \non \\
 &=&  - {N \over 2\pi\Delta} \int_{-\infty}^{\infty}
 d\omega\ 
 {\cos^{2}(\omega \Lambda)\ \sinh((\nu-1)\omega/2) \over 
 \sinh(\nu \omega/2) \cosh(\omega/2)} \,,
\label{SGbulkenergy}
\ee
and
\be
E_{boundary} &=& - {1 \over \Delta} 
\int_{-\infty}^{\infty}d\lambda\ a_{1}(\Lambda - \lambda) 
\left[ R(\lambda) - \delta(\lambda) \right] \non \\
 &=& - {1\over 2\pi\Delta} \int_{-\infty}^{\infty}
 d\omega\ {\cos(\omega \Lambda)\ \sinh((\nu-1)\omega/2) \over 
 \sinh(\nu \omega/2)}
 \left[ {\hat R}(\omega) - 1 \right]\,.
\label{SGboundenergy}
\ee
Taking the continuum limit $N \rightarrow \infty$, $\Delta \rightarrow
0$, $\Lambda \rightarrow \infty$, keeping the length $R$ and the
soliton mass $m$ fixed according to (\ref{continuumlimit}),
we obtain (closing the integrals in the upper half plane and keeping 
only the contribution from the pole at $\omega=i\pi$)
\be
E_{bulk} = {1\over 4} m^{2} R \cot( \nu \pi/2)
\label{SGcontbulkenergy}
\ee
and
\be
E_{boundary} = - {m\over 2} \left[ -\cot(\nu \pi/4) -1 + 
{\cos((\nu -  2 s_{+} a_{+})\pi/2)\over \sin(\nu \pi/2)} +
{\cosh(\pi b_{+})\over \sin(\nu \pi/2)}  + (+ \leftrightarrow -) 
\right]
\,,
\label{SGcontboundenergy}
\ee
where $s_{\pm} = \sgn(a_{\pm} - {1\over 2})$.

The same result (\ref{SGcontbulkenergy}) for the bulk energy was
obtained by a TBA analysis in \cite{LMSS}.  Using the relation
(\ref{bulkparamreltn}) between the lattice and continuum bulk coupling
constants, one arrives at the well-known result \cite{AZ2} for the
bulk energy of the continuum sine-Gordon model.

The result (\ref{SGcontboundenergy}) for the boundary energy, 
which is the sum of contributions from both boundaries, implies 
(see the corresponding discussion for the XXZ chain at the end 
of Section \ref{subsec:XXZboundenergy}) that the
contribution of each boundary is given by
\be
E_{boundary}^{\pm} = - {m\over 2} \left[ -{1\over 2}\cot(\nu \pi/4) 
-{1\over 2} + 
{\cos((\nu -  2 s_{\pm} a_{\pm})\pi/2)\over \sin(\nu \pi/2)} +
{\cosh(\pi b_{\pm})\over \sin(\nu \pi/2)}  \right]
\,.
\ee
Comparing with Al.  Zamolodchikov's result \cite{AZ1, BPT} for the
energy of the continuum sine-Gordon model with a single boundary 
\be
E(\eta \,, \vartheta) = -{m\over 2 \cos \left(\pi/ (2\lambda) \right)}
\left[-{1\over 2}\cos \left(\pi/ (2\lambda) \right) 
+{1\over 2}\sin \left(\pi/ (2\lambda) \right)-{1\over 2} 
+ \cos (\eta/\lambda) + \cosh (\vartheta/\lambda) \right] \,,
\label{AlZam}
\ee 
and using again the bulk relation (\ref{bulkparamreltn}), we conclude
that the boundary parameters of the continuum model ($\eta_{\pm} \,,
\vartheta_{\pm}$) and of the lattice model ($a_{\pm} \,, b_{\pm}$) are
related as follows: \footnote{For simplicity, we assume in the 
remainder of this section that $a_{\pm} > 1/2$, and therefore, 
$s_{\pm} =1$.}
\be
\eta_{\pm} &=& (1 + \lambda - 2\lambda a_{\pm} ){\pi\over 2} 
= (\lambda+1)({\pi\over 2} + i \alpha_{\pm}) \,, \non \\
\vartheta_{\pm} &=& \lambda \pi b_{\pm} = (\lambda+1) \beta_{\pm} \,,
\label{boundparamreltn1}
\ee
where we have also made use of (\ref{newparams}).

Note that the continuum boundary parameters $\eta_{\pm} \,,
\vartheta_{\pm}$ in Al.  Zamolodchikov's result (\ref{AlZam})
are those which appear in the Ghoshal-Zamolodchikov
boundary $S$ matrix \cite{GZ}.  Their relation to the parameters
$\mu_{\pm} \,, \varphi_{0}^{\pm}$ in the boundary action
(\ref{SGboundaction}) is given by \cite{AZ1, BPT}
\be
\cos \left({\beta^{2}\over 8\pi}(\eta_{\pm} + i \vartheta_{\pm}) \right)
&=& {\mu_{\pm}\over \mu_{c}}  
e^{- {i\over 2}\beta \varphi_{0}^{\pm}} \,, \non \\
\cos \left({\beta^{2}\over 8\pi}(\eta_{\pm} - i \vartheta_{\pm}) \right)
&=& {\mu_{\pm}\over \mu_{c}} e^{+ {i\over 2}\beta \varphi_{0}^{\pm}} \,,
\ee
where
\be
\mu_{c} = \sqrt{2 \mu_{bulk}\over
\sin \left({\beta^{2}\over 8\pi}\right)} \,.
\ee 
It follows from (\ref{boundparamreltn1}) that the relation between 
the boundary parameters of the lattice model 
($\alpha_{\pm} \,, \beta_{\pm}$)  and the boundary 
parameters in the continuum action ($\mu_{\pm} \,, 
\varphi_{0}^{\pm}$) is given by
\be
\sinh (\alpha_{\pm} + \beta_{\pm})
&=& {\mu_{\pm}\over \mu_{c}}
i e^{- {i\over 2}\beta \varphi_{0}^{\pm}} \,, \non \\
\sinh (\alpha_{\pm} - \beta_{\pm})
&=& {\mu_{\pm}\over \mu_{c}} 
i e^{+ {i\over 2}\beta \varphi_{0}^{\pm}} \,.
\label{boundparamreltn2}
\ee
For later convenience, we remark that for the left-right symmetric
boundary case with $a_{+}= a_{-}$ and $b_{+} = -b_{-}$, these
relations imply
\be
\varphi_{0}^{+} = -\varphi_{0}^{-} = {1\over \beta} q_{2 
a_{+}}(b_{+}) \,, \qquad 
\mu_{+} = \mu_{-} = \mu_{c} | \sinh \mu( b_{+} + i a_{+}) | \,,
\label{specialrltn}
\ee
where the function $q_{n}(\lambda)$ is defined in (\ref{logfuncts}).

We still have not discussed the relation between the lattice 
parameters $\theta_{\pm}$ and the continuum parameters 
$\gamma_{\pm}$. We conjecture that these boundary parameters 
are related as follows:
\be
\gamma_{\pm} = -\lambda \pi c_{\pm} = i(\lambda+1) \theta_{\pm}
\,.
\label{thetareltn}
\ee
We perform a check on this conjecture at the free-Fermion point
$(\lambda=1$) in Section \ref{subsec:SGcasimirenergy}.  The
constraints on the lattice parameters (\ref{realconstraints}) with
$k=1$ then imply corresponding constraints on the continuum parameters
\be
\eta_{-} + \eta_{+} &=& \mp |\gamma_{-} - \gamma_{+}| + 
\pi  
\,, \non \\
\vartheta_{-} + \vartheta_{+} &=& 0 \,.
\label{SGconstraints}
\ee 

Finally, let us verify that the first relation in
(\ref{boundparamreltn1}) is correct in the Dirichlet limit.  Indeed,
in terms of the Ghoshal-Zamolodchikov boundary parameters $(\xi_{\pm}
\,, k_{\pm})$, which are related to the parameters $(\eta_{\pm} \,,
\vartheta_{\pm})$ by \cite{GZ}
\be
\cos \eta_{\pm} \cosh \vartheta_{\pm} = -{1\over k_{\pm}} \cos 
\xi_{\pm} \,, \qquad 
\cos^{2} \eta_{\pm} + \cosh^{2} \vartheta_{\pm} = 
1 + {1\over k_{\pm}^{2}} \,,
\label{GZparams}
\ee
the Dirichlet limit corresponds to $k_{\pm} \rightarrow 0$, which 
implies $\xi_{\pm} = \eta_{\pm}$.  On the other hand, for this case,
the following relation between lattice and continuum parameters is 
known \cite{FS, DN}: 
$\xi_{\pm} =\left({\nu - 2 a_{\pm}\over \nu - 1} \right){\pi\over 2}
= (1 + \lambda - 2\lambda a_{\pm}){\pi\over 2}$. This result is 
evidently consistent with (\ref{boundparamreltn1}).

\subsection{Nonlinear integral equation and Casimir energy}
\label{subsec:SGcasimirenergy}

We consider now the computation of the Casimir (order $1/R$) energy. 
Rather than follow the Euler-Maclaurin/Wiener-Hopf approach of Section
\ref{subsec:XXZcasimirenergy}, we use instead an approach \cite{KP,
DDV2} based on a nonlinear integral equation for the ground-state
counting function (\ref{SGcounting}), which is of interest in its own
right. 

The derivation of the nonlinear integral equation for the case at hand
is similar to the case of Dirichlet boundary conditions treated in
\cite{LMSS}. Indeed, following the usual steps, we obtain
\be
{d\over d\lambda} h(\lambda) &=& N \left[ 
s(\lambda + \Lambda) + s(\lambda - \Lambda) \right] + R(\lambda) 
-\int_{-\infty}^{\infty}{d\lambda'\over 2\pi i} 
G(\lambda-\lambda' + i \epsilon)
{d\over d\lambda'} \ln (1 - e^{-2\pi i h(\lambda'- i \epsilon)}) \non \\
&+& \int_{-\infty}^{\infty}{d\lambda'\over 2\pi i} 
G(\lambda-\lambda' - i \epsilon)
{d\over d\lambda'} \ln (1 - e^{2\pi i h(\lambda'+ i \epsilon)}) \,,
\label{deriv}
\ee
where $\epsilon$ is a small positive quantity, ${\hat G}(\omega) = 
{\hat a}_{2}(\omega)/\left(1+{\hat a}_{2}(\omega)\right)$, and 
$s(\lambda)$ and $R(\lambda)$ are given by (\ref{kernels}).
Moreover, the energy  is given by
\be
E &=& E_{bulk} + E_{boundary} -
{1\over \Delta}\int_{-\infty}^{\infty}{d\lambda\over 2\pi i} 
s'(\Lambda-\lambda + i \epsilon)
\ln (1 - e^{-2\pi i h(\lambda- i \epsilon)}) \non \\
&+& {1\over \Delta}\int_{-\infty}^{\infty}{d\lambda\over 2\pi i} 
s'(\Lambda-\lambda - i \epsilon)
\ln (1 - e^{2\pi i h(\lambda+ i \epsilon)}) \,,
\ee
where $E_{bulk}$ and $E_{boundary}$ are given by (\ref{SGbulkenergy})
and (\ref{SGboundenergy}), respectively; and a prime on a function
denotes differentiation with respect to its argument.  Integrating
(\ref{deriv}), taking the continuum limit as before
(\ref{continuumlimit}), changing to the rescaled rapidity variable
$\theta \equiv \pi \lambda$, and setting $f(\theta) \equiv 2\pi i
h(\theta)$, we finally obtain the nonlinear integral equation
\be
f(\theta) = 2i m R \sinh \theta + i P_{bdry}(\theta) 
+ {2i\over \pi} \int_{-\infty}^{\infty} d\theta'\ \Im m\
G(\theta-\theta' - i \epsilon)\
\ln (1 - e^{f(\theta' + i \epsilon)}) \,,
\label{DDVeqn}
\ee
where 
\be
G(\theta) =  {1\over 2\pi} \int_{-\infty}^{\infty} d\omega\ 
e^{-i \omega \theta/\pi} {\sinh((\nu-2)\omega/2) \over 
2\sinh((\nu-1)\omega/2) \cosh(\omega/2)} \,,
\label{Gtheta}
\ee 
and $P_{bdry}(\theta)$ is the odd function satisfying
$P_{bdry}'(\theta) = 2 R(\theta)$. Moreover,
\be
E = E_{bulk} + E_{boundary} + E_{Casimir} \,,
\ee
where $E_{bulk}$ and $E_{boundary}$ are now given by
(\ref{SGcontbulkenergy}) and (\ref{SGcontboundenergy}), respectively;
and $E_{Casimir}$ is given by
\be
E_{Casimir} = -{m\over 2\pi} \int_{-\infty}^{\infty} d\theta\ 
\Im m\ 
\sinh (\theta+ i \epsilon)  \ln (1 - e^{f(\theta + i \epsilon)})
\,.
\label{SGCasimir}
\ee

\subsubsection{Ultraviolet limit}

Let us now consider the ultraviolet limit $R \rightarrow 0$.
Proceeding as in \cite{LMSS}, we obtain 
$E_{Casimir} = -  c \pi/(24 R)$, where the central charge $c$ is given by 
\be
c=1 - {6\over \pi^{2}}\left({\nu-1\over \nu}\right)(\sigma-\pi)^{2}
\,, \label{SGcc1}
\ee
and
\be
\sigma = P_{bdry}(\infty) = \pi \left\{
1 + {1\over \nu-1}\left[ {\nu\over 2}\left(s_{+} + s_{-}\right)
-\left(a_{-}+a_{+}-ib_{-}-ib_{+}-1 \right) \right] \right\}
\,, \label{SGcc2}
\ee
where $s_{\pm} = \sgn(a_{\pm} - {1\over 2})$.
We conclude that the value of the central charge for the sine-Gordon
model coincides with the result (\ref{centralcharge1}),
(\ref{centralcharge2}) for the XXZ spin chain. In terms of the 
sine-Gordon parameters $\gamma_{\pm}$ (\ref{thetareltn}), the central 
charge is given by
\be
c = 1 - {6 \over \nu (\nu -1)} 
\left[ {\nu\over 2}\left(s_{+} + s_{-}\right)
\mp {(\nu-1)\over \pi}|\gamma_{+} - \gamma_{-}| 
\right]^{2}\,.
\ee

\subsubsection{Free-Fermion point}

A dramatic simplification occurs at the free-Fermion point $\beta^{2}
= 4\pi$, which corresponds (\ref{bulkparamreltn}) to $\lambda=1$, or
$\nu = 2$.  Indeed, for this value of the bulk coupling constant, the
kernel $G(\theta)$ (\ref{Gtheta}) vanishes.  It immediately follows
from (\ref{DDVeqn}) that $f(\theta)$ is given by
\be
f(\theta) = 2i m R \sinh \theta + i P_{bdry}(\theta) \,.
\label{fFFpoint}
\ee
Let us now rewrite the expression (\ref{SGCasimir}) for the Casimir
energy as
\be
E_{Casimir} = -{m\over 4\pi i} \int_{-\infty}^{\infty} d\theta\ 
\sinh \theta\ \Big\{ \ln (1 - e^{f(\theta + i \epsilon)})
- \ln (1 - e^{-f(\theta - i \epsilon)}) \Big\}
\,;
\ee
and then change integration variables $\theta' = \theta - {i\pi\over
2} + i \epsilon$ in the first integral, and $\theta' = \theta +
{i\pi\over 2}- i \epsilon$ in the second integral.  Assuming that the
resulting contours can then be deformed to the real axis, and dropping
the primes, we obtain
\be
E_{Casimir} &=& -{m\over 4\pi} \int_{-\infty}^{\infty} d\theta\ 
\cosh \theta\ \left\{ \ln \left(1 - e^{f(\theta + {i\pi\over 2})}\right)
+ \ln \left(1 - e^{-f(\theta - {i\pi\over 2})}\right) \right\} \non \\
&=& -{m\over 4\pi} \int_{-\infty}^{\infty} d\theta\ 
\cosh \theta\  \ln \left(1 - e^{f(\theta + {i\pi\over 2})}\right)
\left(1 - e^{-f(\theta - {i\pi\over 2})}\right)  \,.
\label{SGCasimir2}
\ee 
Using (\ref{fFFpoint}), we obtain
\be
E_{Casimir} =-{m\over 2\pi} \int_{0}^{\infty} d\theta\ 
\cosh \theta\  \ln \left( 1 + E_{1}(\theta)\ e^{-2m R \cosh \theta}
+ E_{2}(\theta)\ e^{-4m R \cosh \theta} \right) \,,
\label{SGCasimirFF}
\ee
where
\be
E_{1}(\theta) = -e^{i  P_{bdry}(\theta + {i\pi\over 2})}
- e^{-i P_{bdry}(\theta - {i\pi\over 2})} \,, \qquad
E_{2}(\theta) = e^{i P_{bdry}(\theta + {i\pi\over 2})}
e^{-i P_{bdry}(\theta - {i\pi\over 2})} \,.
\label{E1E2}
\ee
One can show using (\ref{kernels}) and (\ref{boundparamreltn1}) that
$e^{i P_{bdry}(\theta + {i\pi\over 2})}$ is given (for $\lambda=1$) by
\be
e^{i P_{bdry}(\theta + {i\pi\over 2})} = 
{\sinh((\theta +  i \eta_{+})/2)\over
\cosh((\theta - i\eta_{+})/2)}
{\sinh((\theta +  i\eta_{-})/2)\over
\cosh((\theta - i\eta_{-})/2)}
{\sinh((\theta - \vartheta_{+})/2)\over
\cosh((\theta + \vartheta_{+})/2)}
{\sinh((\theta - \vartheta_{-})/2)\over
\cosh((\theta + \vartheta_{-})/2)} \,, 
\ee
and $e^{-i P_{bdry}(\theta - {i\pi\over 2})}$ is given by the 
complex conjugate of the above expression. 

This result can now be compared with the result obtained using the TBA
approach of Caux et al.  \cite{CSS2}.  One finds that the Casimir
energy is again given by (\ref{SGCasimirFF}), with (see Eq.  (58) in
\cite{CSS2})
\be
E_{1}(\theta) = \tr \left( {\bar K}_{-}(\theta)\ K_{+}(\theta) \right) 
\,, \qquad
E_{2}(\theta) = \det  \left( {\bar K}_{-}(\theta)\ K_{+}(\theta) \right) 
\,, \label{E1E2CSS}
\ee
where $K_{\pm}(\theta)$ are the crossed-channel boundary $S$ matrices
\cite{GZ}
\be
K_{\pm}(\theta) = r_{\pm}({i \pi\over 2} - \theta)
\left( \begin{array}{cc}
- {i k_{\pm}\over 2} e^{-i\gamma_{\pm}} \sinh 2\theta
& \sin (\xi_{\pm} - i\theta) \\
 -\sin (\xi_{\pm} + i\theta) 
& - {i k_{\pm}\over 2} e^{i\gamma_{\pm}} \sinh 2\theta
\end{array} \right) \,.
\label{boundSmatrix}
\ee
Note that we have included in the boundary $S$ matrices their
dependence on the (real) parameters $\gamma_{\pm}$,
corresponding to the ${d\varphi\over dy}$ terms in the boundary action
(\ref{SGboundaction}). \footnote{The relation between the 
$\gamma_{\pm}$ parameters in the boundary $S$ matrix
(\ref{boundSmatrix}) and those in the boundary action
(\ref{SGboundaction}) is not {\it a priori} obvious.  The fact that
these parameters are the same (and, in particular, that the 
normalization of the ${d\varphi\over dy}$ terms  in the boundary 
action is correct) follows from the observation \cite{GZ} that
a shift $\gamma_{+} \mapsto  \gamma_{+} + \gamma$ 
in the boundary $S$ matrix implies a corresponding shift
$B_{+}(\varphi \,, {d\varphi\over dy}) \mapsto 
B_{+}(\varphi \,, {d\varphi\over dy}) +  
{\pi\gamma\over \beta} {d\varphi\over dy}$ in the boundary 
action.}
The scalar factors $r_{\pm}(\theta)$ are given by
\be
r_{\pm}(\theta) = {1\over \cos \xi_{\pm}} 
\sigma(\eta_{\pm} \,, -i\theta)\ \sigma(i\vartheta_{\pm} \,, -i\theta)
\,,
\ee
where \cite{AKL}
\be
\sigma(x \,, u) = {\cos x\over 2 
\cos\left( {\pi\over 4} + {x\over 2} -{u\over 2} \right)
\cos\left( {\pi\over 4} - {x\over 2} -{u\over 2} \right)} \,.
\ee
Using the relations (\ref{GZparams}) to express $K_{\pm}(\theta)$ in
terms of the boundary parameters $\eta_{\pm} \,, \vartheta_{\pm} \,,
\gamma_{\pm}$, we find that the results (\ref{E1E2}),
(\ref{E1E2CSS}) for $E_{1}(\theta)$ and $E_{2}(\theta)$ agree when the
boundary parameters satisfy the constraints (\ref{SGconstraints}). 
This is a good check on our results (\ref{DDVeqn}),
(\ref{SGCasimir}) for the Casimir energy for general values of the
bulk coupling constant, as well as on the conjectured relation
(\ref{thetareltn}) between the boundary parameters $\theta_{\pm}$ and
$\gamma_{\pm}$.

\subsubsection{General values of $R$ and $\nu$}

\begin{figure}[tb]
	\centering
	\includegraphics[width=0.80\textwidth]{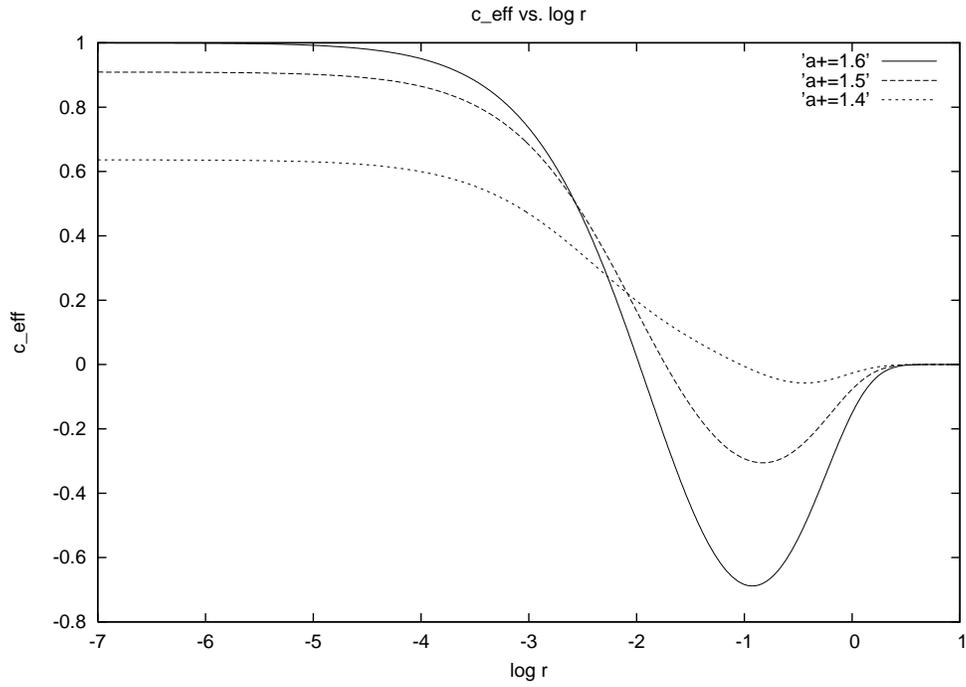}
	\caption[xxx]{\parbox[t]{1.0\textwidth}{
	$c_{eff}$ vs. $\log r$, for $a_{+}=a_{-}= 1.4\,, 1.5\,, 1.6$, 
	with $\nu=2.2$ and $b_{+}=-b_{-}=1.3$}
	}
	\label{fig:graph1}
\end{figure}

\begin{figure}[htb]
	\centering
	\includegraphics[width=0.80\textwidth]{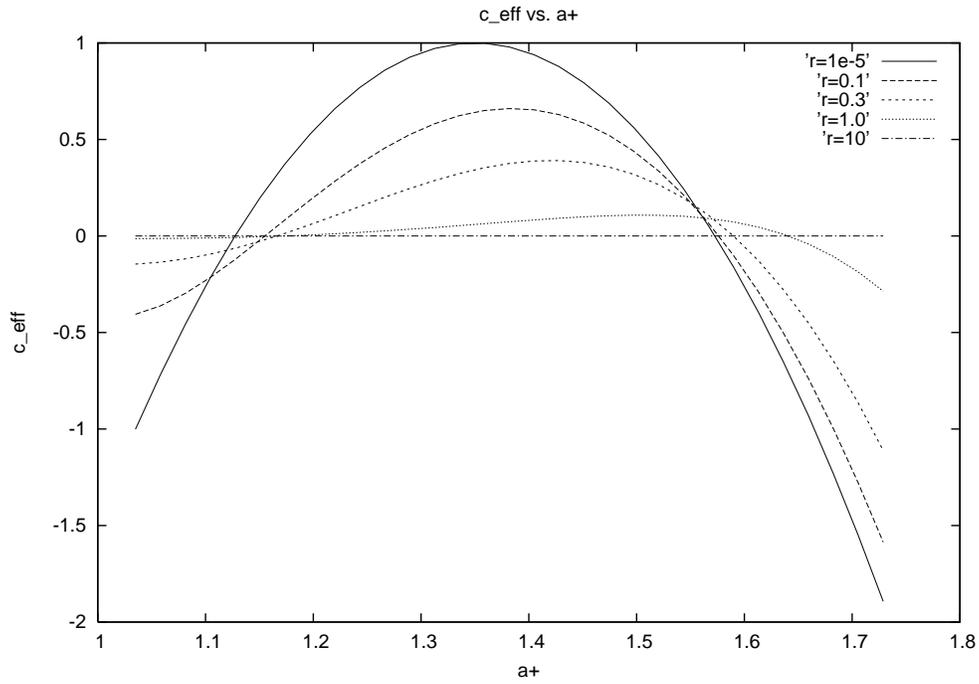}
	\caption[xxx]{\parbox[t]{1.0\textwidth}{
	$c_{eff}$ vs. $a_{+}=a_{-}$, for $r= 10^{-5}\,, 0.1\,, 0.3\,,
	1.0\,, 10$, with $\nu=1.7$ and $b_{+}= b_{-}=0$}
	}
	\label{fig:graph2}
\end{figure}

\begin{figure}[htb]
	\centering
	\includegraphics[width=0.80\textwidth]{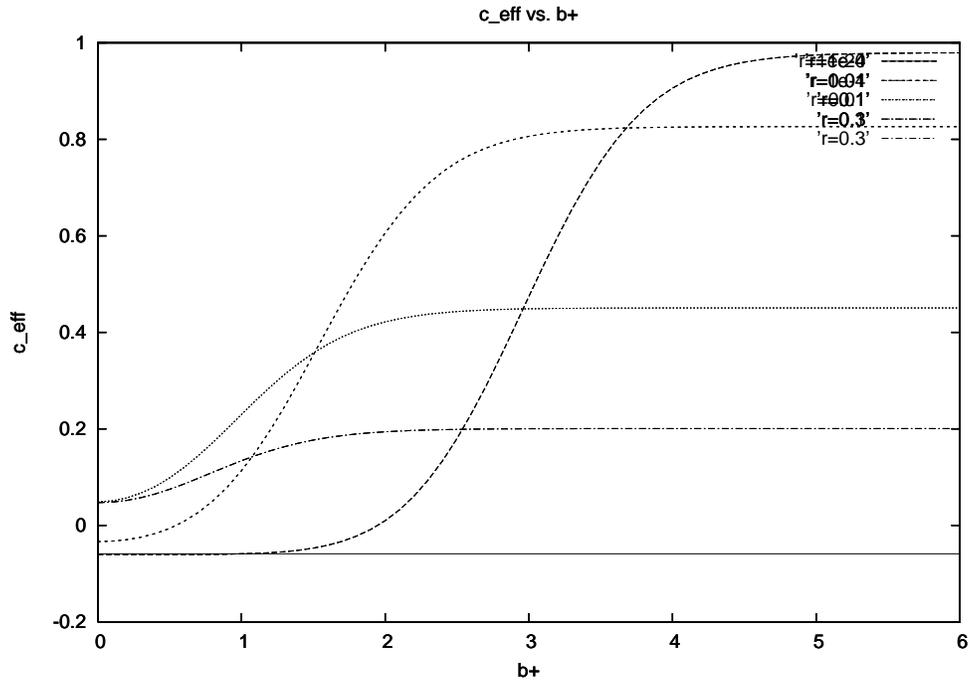}
	\caption[xxx]{\parbox[t]{1.0\textwidth}{
	$c_{eff}$ vs. $b_{+}=-b_{-}$, for $r= 10^{-20}\,, 10^{-4}\,,
	0.01\,, 0.1\,, 0.3$, with $\nu=2.7$ and $a_{+}=a_{-}=1.4$}
	}
	\label{fig:graph3}
\end{figure}

\begin{figure}[htb]
	\centering
	\includegraphics[width=0.80\textwidth]{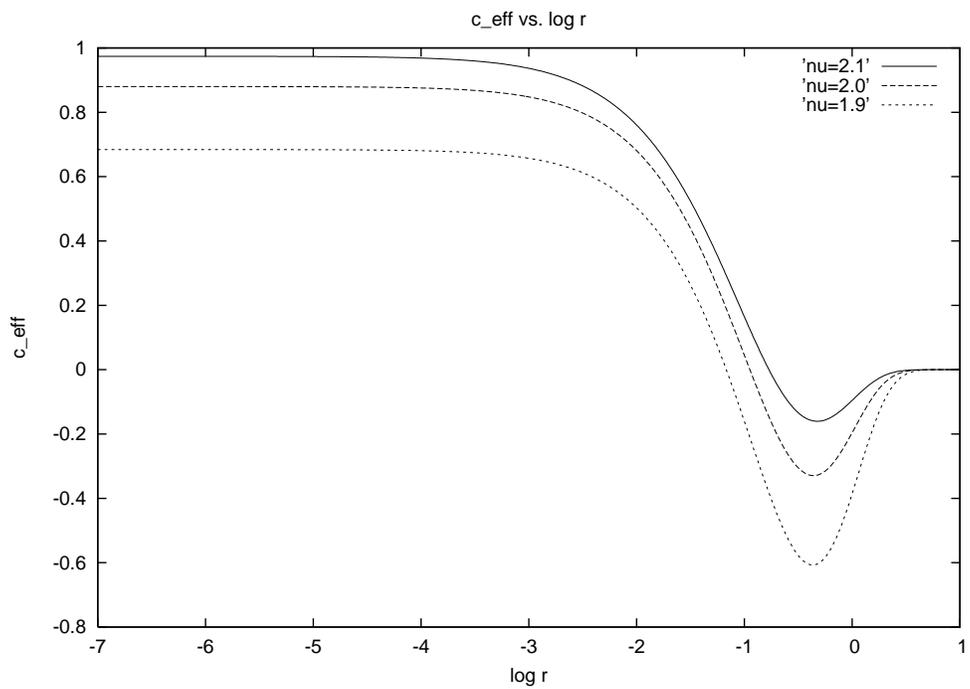}
	\caption[xxx]{\parbox[t]{1.0\textwidth}{
	$c_{eff}$ vs. $\log r$, for $\nu=1.9\,, 2.0\,, 2.1$,
	with $a_{+}=1.7\,, a_{-}= 1.5\,, b_{+}=-b_{-}=0.5$}
	}
	\label{fig:graph4}
\end{figure}

For general values of the length $R$ and of the bulk coupling constant
$\nu$, the Casimir energy cannot be computed analytically. 
Nevertheless, one can readily solve the nonlinear integral equation
(\ref{DDVeqn}) by iteration and compute the Casimir energy numerically
through (\ref{SGCasimir}).\footnote{A useful trick \cite{Ra} is to
consider (\ref{DDVeqn}) with the shift $\theta \mapsto \theta + i
\epsilon$, and to work in a range of $\epsilon$ (typically, centered
at $\epsilon \sim 0.3$) for which the Casimir energy does not depend
on the particular $\epsilon$ value.} Sample results are summarized in
Figures \ref{fig:graph1} - \ref{fig:graph4}, which show the dependence
of $c_{eff} \equiv -24 R E_{Casimir}/\pi$ on the various parameters. 
Note that $r \equiv m R$.  In all cases, the computed value of
$c_{eff}$ in the ultraviolet region $r \rightarrow 0$ agrees with the
analytical result (\ref{SGcc1}), (\ref{SGcc2}).  Also, as expected,
$c_{eff} \rightarrow 0$ in the infrared region $r \rightarrow \infty$. 
Moreover, one can observe the crossover in $c_{eff}$ from the
ultraviolet to the infrared regions.

These graphs are parametrized in part by the boundary parameters
$a_{\pm}\,, b_{\pm}$, in terms of which the function $R(\theta)$ is
defined (\ref{kernels}).  Nevertheless, it is straightforward to
translate to the sine-Gordon boundary parameters using
(\ref{newparams}), (\ref{boundparamreltn2}) - (\ref{thetareltn}). 
Indeed, consider Figure \ref{fig:graph2}, for which $a_{+} = a_{-}$
and $b_{\pm}=0$; and therefore (\ref{realconstraints}), $|c_{-}-c_{+}| =
|2 a_{+}-1|$.  This corresponds to $\varphi_{0}^{\pm}=0$ and
$\mu_{+} = \mu_{-} = \mu_{c} \sin (\pi a_{+}/\nu)$; and also
$|\gamma_{-} - \gamma_{+}| =  \pi |2 a_{+}-1|/(\nu-1)$.
Hence, one can infer from this graph the dependence of $c_{eff}$
on $\mu_{+} = \mu_{-}$ or $|\gamma_{-} - \gamma_{+}|$, keeping
$\varphi_{0}^{\pm}$ fixed.
Similarly, for Figure \ref{fig:graph3}, $a_{+} = a_{-}=1.4$ and $b_{+}
= -b_{-}$, which implies $\varphi_{0}^{+}= -\varphi_{0}^{-} = {1\over
\beta} q_{2.8}(b_{+})$ and $\mu_{+} = \mu_{-} = \mu_{c} | \sinh \mu(
b_{+} +  i 1.4)|$.  Hence, one can infer from this graph the dependence
of $c_{eff}$ on $\mu_{+} = \mu_{-}$ or $\varphi_{0}^{+}=
-\varphi_{0}^{-}$, keeping $|\gamma_{-} - \gamma_{+}|$ fixed.

Finally, we remark that the convergence of the iterative procedure
which we use to numerically solve (\ref{DDVeqn}) depends sensitively
on the values of the various parameters.  Owing to the great number of
parameters, we have not attempted to find the entire domain of
convergence.

\section{Discussion}\label{sec:conclude}

We have exploited the recently-proposed \cite{Ne1, CLSW, Ne2} Bethe
Ansatz solution of the open XXZ chain with nondiagonal boundary terms
to compute finite size effects in both the XXZ and sine-Gordon models,
in a range of parameter space previously not possible.  Although we
have focused here exclusively on properties of the ground state, it
should be possible, and quite interesting, to generalize this work to
excited states, with bulk and/or boundary excitations. Such a study 
has recently been made for the case of Dirichlet boundary 
conditions \cite{ABR}. It would also be interesting to introduce a 
``twist'' in the nonlinear integral equation to study $\phi_{13}$
perturbed minimal models with boundaries, and to consider applications 
of our results to condensed-matter systems. 

It would be desirable to investigate these models for the full range
of boundary parameters, unhampered by the constraint
(\ref{constraint}).  Indeed, this constraint precludes an
investigation of the Casimir energy of the sine-Gordon model 
in the massless scaling limit as a
function of $\chi \equiv {\beta\over 2}(\varphi_{0}^{+} -
\varphi_{0}^{-})$, which is of interest in certain condensed-matter
applications \cite{CSS1, LR, CSS2}.  However, finding a Bethe Ansatz
solution for this most general case remains a challenging open
problem.

\section*{Acknowledgments}

We are grateful to F. Ravanini for providing us with sample code for
numerical solution of the nonlinear integral equation; and to O.
Alvarez for his help in preparing a figure.  This work was supported in
part by the Korea Research Foundation 2002-070-C00025 (C.A.); and by
the National Science Foundation under Grants PHY-0098088 and
PHY-0244261, and by a UM Provost Award (R.N.).

\eject 

\end{document}